\documentclass[final,3p,times]{elsarticle}

\usepackage{amssymb}
\usepackage{amsmath}
\usepackage{graphicx}
\usepackage{longtable}
\usepackage{lineno}
\usepackage{multirow}
\usepackage{threeparttable}
\usepackage{booktabs}
\usepackage{tabularx}
\usepackage{float}
\usepackage{placeins}
\usepackage{siunitx}
\usepackage{hyperref}

\hypersetup{
	colorlinks = true,
	linkcolor  = blue,
	citecolor  = blue,
	urlcolor   = blue,
	hypertexnames = false
}
\usepackage{doi}
\graphicspath{{figures/}}

\biboptions{numbers,sort&compress}

\begin{document}
	
	\begin{frontmatter}

	\title{Spatiotemporal organization in bike-sharing systems using gravity model parameters}
	
	\author[aff1,aff2,aff3]{Hanbo Zhang}
	\author[aff1,aff2,aff3]{Qi Rao}
	\author[aff1,aff2,aff3]{Bo Yang\corref{cor1}}
	\cortext[cor1]{Corresponding author}
	\ead{yangbo@kust.edu.cn}

	\affiliation[aff1]{organization={Data Science Research Center, Kunming University of Science and Technology},
		addressline={727 South Jingming Road},
		city={Kunming},
		postcode={650500},
		country={China}}

	\affiliation[aff2]{organization={Faculty of Science, Kunming University of Science and Technology},
		addressline={727 South Jingming Road},
		city={Kunming},
		postcode={650500},
		country={China}}

	\affiliation[aff3]{organization={Yunnan Key Laboratory of Complex Systems and Brain-Inspired Intelligence, Kunming University of Science and Technology},
		addressline={727 South Jingming Road},
		city={Kunming},
		postcode={650500},
		country={China}}

\begin{abstract}
	Gravity models describe bike-sharing origin--destination (OD) flows as increasing with origin and destination activity and decreasing with distance. Yet how these relationships change within a day and across observation areas remains unclear. We examine temporal and spatial variation in origin, destination and distance exponents \(\alpha,\beta,\gamma\), together with \(R^2\) and mean absolute error (MAE), across eight bike-sharing systems. Temporal modeling uses sliding windows, while spatial modeling expands a circular area and separates intra-zonal, cross-zonal outflow, cross-zonal inflow and extra-zonal trips. We find recurring patterns across cities: morning and evening peaks differ in origin and destination dependence, while cross-zonal outflow and inflow across the same boundary show opposite changes in their relative origin and destination dependence as radius increases. Model performance and distance dependence also vary with time and radius. A full-day, citywide fit therefore combines time periods and flow types with different gravity relationships.
\end{abstract}


\end{frontmatter}

\label{sec:intro}

Human mobility exhibits regular and reproducible patterns across space and time \cite{Zhong2025Universal,Tan2025Spatiotemporal,Pappalardo2023FutureDirections}, and these movements shape urban spatial structure \cite{Xu2021UrbanGrowth}, traffic dynamics \cite{Zhao2024}, and the spatial distribution of human activity \cite{Ramani2024WFH}. Shared mobility has become increasingly important in urban transport because of its potential to reduce energy use and emissions \cite{Winkler2023SustainableMobility} and support sustainable travel \cite{Asensio2022Micromobility}. Bike-sharing is a form of shared mobility that reflects human movement between urban locations, with trips represented as origin--destination (OD) flows \cite{Liang2024SharedMobilityOD}. These flows exhibit clear morning and evening peaks \cite{Xu2023JAG,Waldner2025BikeSharing} and are unevenly distributed across urban space \cite{Meng2023,LiWang2025SharedBicycle}. Uncovering these temporal and spatial patterns can help explain how short-distance mobility is organized within cities, provide insight into urban functioning \cite{Tan2025Spatiotemporal}, and support urban planning \cite{Abbiasov2024FifteenMinute}, environmental management \cite{Winkler2023SustainableMobility}, and traffic prediction \cite{Liang2024SharedMobilityOD,Lu2025BikeSharing}.

Numerous models have been developed to describe human mobility flows between locations \cite{Barbosa2018}. One of the earliest and most widely used is the gravity model, which assumes that flow increases with the populations at the origin and destination and decreases with distance \cite{Zipf1946,Wilson1967TR,Evans1973TR}. The radiation model also considers opportunities located nearer to the origin than the selected destination, while the universal opportunity model describes exploratory and cautious tendencies when travelers compare possible destinations \cite{Simini2012,LiuYan2020}. Deep Gravity retains the general form of the gravity model but uses a deep neural network to learn the combined effects of multiple urban factors \cite{Simini2021}. More recently, neuroGravity combined the gravity law with graph neural networks to reconstruct mobility networks from limited observations and transfer learned relationships to cities without observed flows \cite{Yang2026NeuroGravity}. Despite these developments, the gravity model remains valuable for analyzing highway flows \cite{Jung2008}, subway flows \cite{Goh2012}, bus flows \cite{HongJung2016}, airline flows \cite{Grosche2007}, and urban mobility flows \cite{Masucci2013,Kwon2023MultipleGravity,Thompson2019,ZhangX2024}. Its simple and interpretable form makes it particularly suitable for examining how the dependence of mobility flows on origin activity, destination activity, and travel distance varies across temporal and spatial settings \cite{CabanasTirapu2025GravityLike}. For bike-sharing systems, Li et al. showed that the gravity model can predict dockless bike-sharing flows at fine spatial resolutions and that its population and distance exponents change with spatial resolution \cite{LiR2021}. Jouve et al. further showed that gravity model parameters capture weekday--weekend and seasonal differences in bike-sharing behavior, while changes in distance-related behavior can persist after COVID-19 restrictions \cite{Jouve2025}.

However, human activity varies markedly within a day and across different parts of a city \cite{Tan2025Spatiotemporal,Santana2023TimeSpace,Boucherie2025Geography}, making finer temporal and spatial analysis necessary for understanding bike-sharing OD flows. Gravity models fitted to trips aggregated over long periods or across an entire study area may fail to capture changes in how flows depend on origin activity, destination activity, and distance. Despite this limitation, few bike-sharing studies have examined how temporal and spatial variation affects gravity model parameters and model performance. Existing temporal work has examined model estimates over multiweek windows rather than within a day, leaving intraday variation unclear \cite{Jouve2025}. Spatial studies have mainly examined how model estimates change with spatial resolution \cite{LiR2021,Jouve2025}, rather than how they change as the observation range is gradually expanded at a fixed resolution. It therefore remains unclear how gravity model parameters and model performance vary with observation radius \(r\), whether these changes differ among flow types, and whether similar patterns occur in docked and dockless systems.

In this study, we separately analyze temporal and spatial variation in gravity model parameters and model performance using data from docked and dockless bike-sharing systems. The exponents \(\alpha\), \(\beta\), and \(\gamma\) measure origin, destination, and distance dependence, respectively, while \(R^2\) and \(\mathrm{MAE}\) measure model performance. For each city and day type, \(O_i\) and \(D_j\) denote the out-strength of origin \(i\) and the in-strength of destination \(j\), respectively, and remain fixed across time windows and observation radii. Temporal modeling uses sliding time windows, while spatial modeling expands a circular observation area at a fixed spatial resolution and separates trips into II, IO, OI, and OO. We find recurring differences between morning and evening peaks and opposite radial changes in the relative origin and destination dependence of IO and OI. Model performance and distance dependence also vary with time and radius. A full-day, citywide fit therefore combines time periods and flow types with different gravity relationships.

\section{Results}
\label{sec:results}

\subsection{Temporal modeling}
\label{sec:results_temporal}

Intraday trip-start and trip-distance distributions are reported in Supplementary Section~S1. To evaluate the regression performance of the gravity model for
bike-sharing OD flows, we compared it with the radiation model
\cite{Simini2012}, the universal opportunity model
\cite{LiuYan2020}, and a gravity model based on resident population estimates for the
corresponding year \cite{LiR2021}. At \(L=1000\)~m, the flow-based gravity model achieved the highest
\(R^2\) and the lowest \(\mathrm{MAE}\) in all eight systems on both
weekdays and weekends (see Supplementary Section~S2). We further compared the regression performance of the flow-based and
population-based gravity models across trip-distance windows on
weekdays. The flow-based gravity model generally had
lower \(\mathrm{MAE}\) and higher local \(R^2\) than the
population-based gravity model, although the differences between the
two models narrowed at longer distances
(see Supplementary Section~S2). The flow-based gravity model also
outperformed the population-based gravity model across all grid-cell
sizes from \(500\) to \(5000\)~m
(see Supplementary Section~S3).

\begin{figure}[h]
	\centering
	\includegraphics[width=\textwidth]{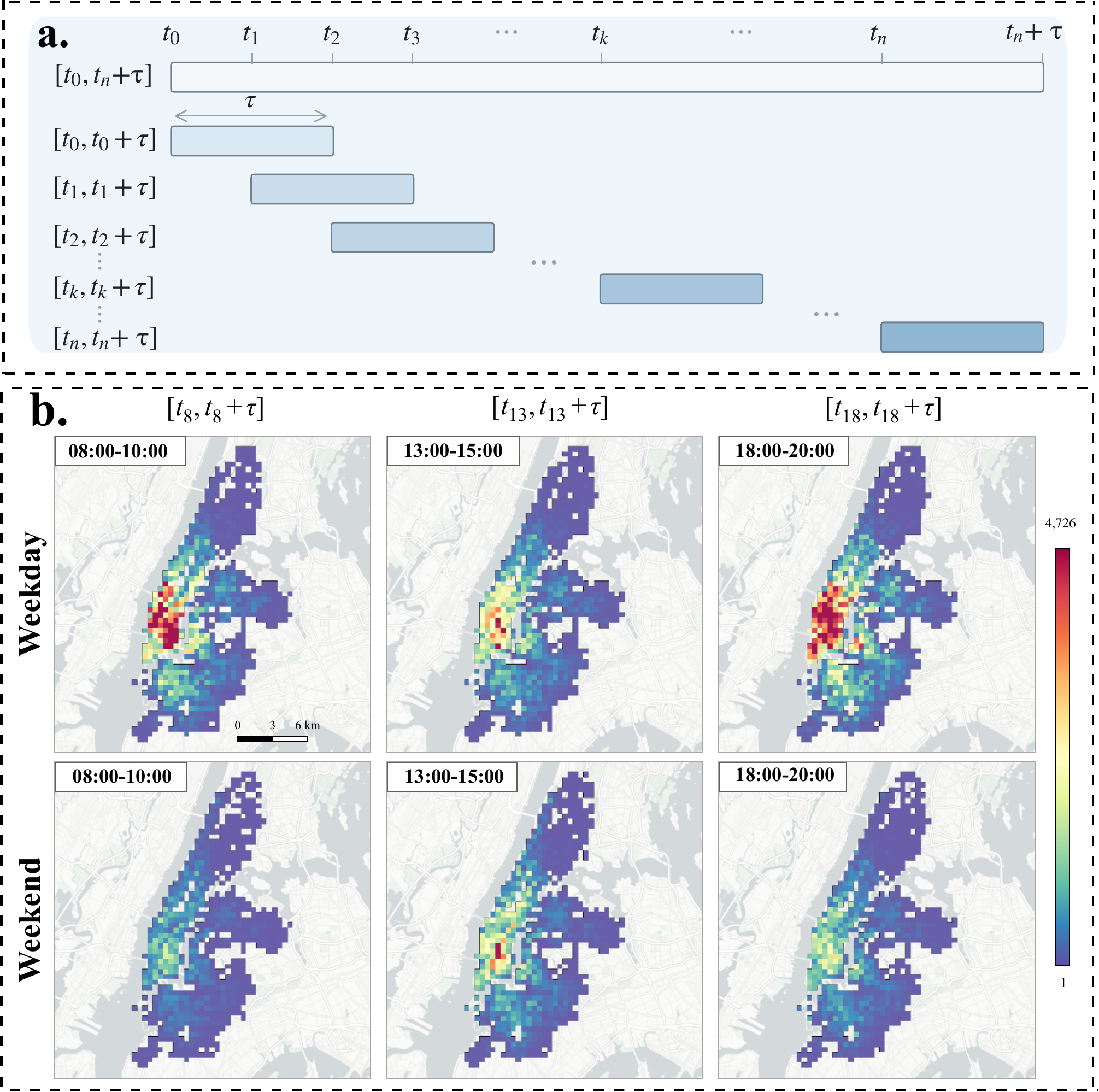}
	\caption{Temporal modeling framework and examples of bike-sharing
		origin trip volume in New York City at \(L=500\)~m.
		(a) Sliding 2~h observation windows with a 1~h step.
		(b) Spatial distributions of cumulative origin trip volume during
		the 08:00--10:00, 13:00--15:00, and 18:00--20:00 windows on
		weekdays and weekends. Grid colors indicate cumulative trip volume.}
	\label{fig:methods_windows}
\end{figure}

\begin{figure}[h]
	\centering
	\includegraphics[width=\textwidth]{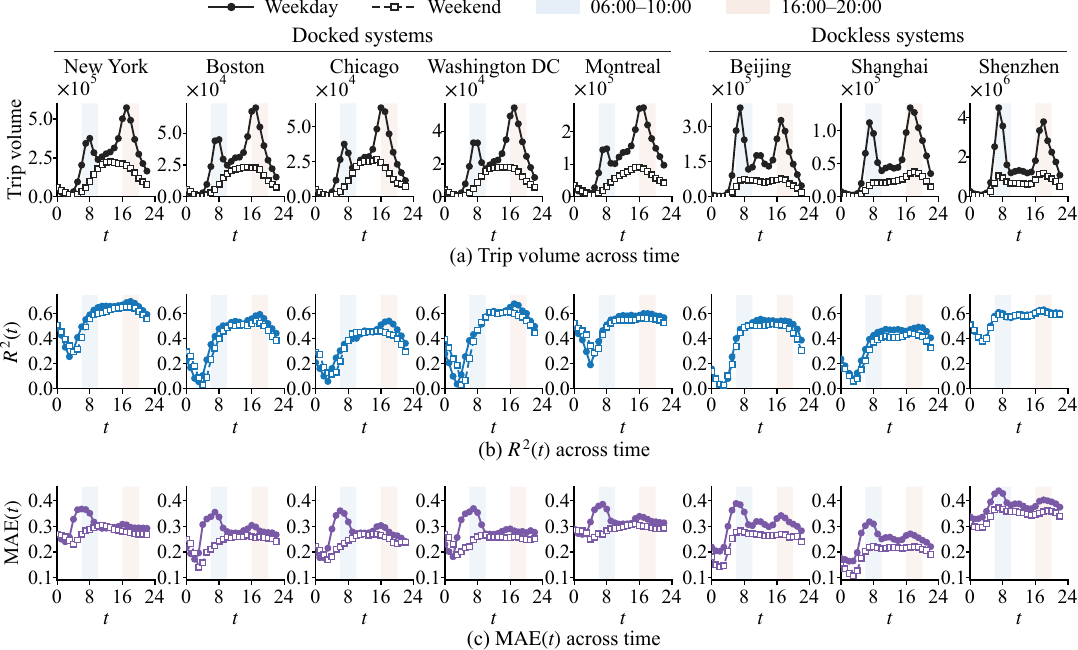}
	\caption{Temporal trip volume and gravity model performance for five docked and three dockless systems, calculated in 2~h observation windows. (a) Trip volume; (b) \(R^2(t)\); (c) \(\mathrm{MAE}(t)\). Solid and dashed curves denote weekdays and weekends, respectively; shaded bands mark 06:00--10:00 and 16:00--20:00.}
	\label{fig:temporal_fit}
\end{figure}

Given the short trip distances reported in Table~\ref{tab:data}, the
temporal analysis uses a grid-cell size of \(L=1000\)~m. The main
analysis uses a 2~h observation window, which retains intraday
variation while providing sufficient positive-flow OD pairs for model
estimation. Figure~\ref{fig:methods_windows} summarizes the sliding-window design
used for the temporal analysis.
Figure~\ref{fig:temporal_fit}
shows that gravity model performance varies within the day and does
not simply follow the trip volume cycle. Weekday trip volume increases rapidly after 06:00 and forms morning and evening peaks (Figure~\ref{fig:temporal_fit}(a)). \(R^2(t)\) rises from its early-morning minimum, remains high during the daytime, and decreases late in the day, with Shenzhen showing the weakest difference between daytime and night (Figure~\ref{fig:temporal_fit}(b)). Weekday \(\mathrm{MAE}(t)\) reaches its maximum during 06:00--10:00 and then decreases, while weekend curves are generally lower and flatter (Figure~\ref{fig:temporal_fit}(c)). After the morning peak, \(R^2(t)\) remains high while \(\mathrm{MAE}(t)\) decreases, showing stronger fit within the corresponding daytime OD flows.

\begin{figure}[H]
	\centering
	\includegraphics[width=0.89\textwidth]{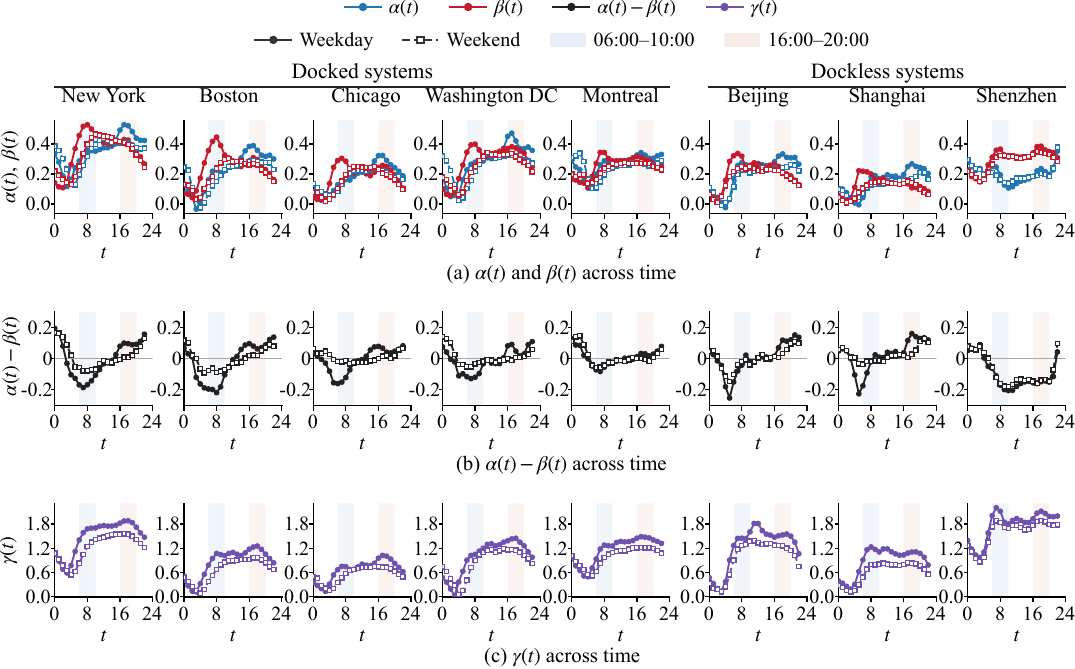}
	\caption{Temporal gravity model parameters for five docked and three dockless systems, estimated separately for 2~h observation windows. (a) Origin and destination dependence exponents, \(\alpha(t)\) and \(\beta(t)\); (b) their difference, \(\alpha(t)-\beta(t)\); (c) the distance dependence exponent, \(\gamma(t)\). Solid and dashed curves denote weekdays and weekends, respectively; shaded bands mark 06:00--10:00 and 16:00--20:00.}
	\label{fig:temporal_parameters}
\end{figure}

Figure~\ref{fig:temporal_parameters} shows clear and systematic intraday changes in the gravity model parameters. On weekday mornings, \(\beta(t)\) exceeds \(\alpha(t)\) in every city, and \(\alpha(t)-\beta(t)\) is therefore negative (Figures~\ref{fig:temporal_parameters}(a) and~\ref{fig:temporal_parameters}(b)). Because origin out-strength and destination in-strength remain fixed across observation windows, the negative difference indicates stronger destination dependence during the morning. Later in the day, \(\alpha(t)-\beta(t)\) increases and becomes positive in every city except Shenzhen, showing a shift from destination dependence to origin dependence during the evening. Shenzhen retains stronger destination dependence through most daytime windows. Weekend \(\alpha(t)-\beta(t)\) curves generally show the same morning
decrease and subsequent recovery, although the magnitude differs among
cities.

The distance dependence exponent \(\gamma(t)\) decreases after midnight, reaches its minimum in the early morning, and increases rapidly after 06:00 (Figure~\ref{fig:temporal_parameters}(c)). It remains relatively high during the daytime and decreases late in the day. Weekday values exceed weekend values during most active hours. Because a larger \(\gamma(t)\) indicates stronger distance dependence, distance dependence is stronger during the daytime than at night and stronger on weekdays than on weekends. Shenzhen shows the highest daytime distance dependence among the eight cities.

The results obtained using the 1~h and 3~h observation windows are similar to those obtained using the 2~h window, indicating that the choice of observation window does not substantially affect the
temporal variations in gravity model performance and parameters
(see Supplementary Section~S4).

\subsection{Spatial modeling}
\label{sec:results_spatial}

\begin{figure}[h]
	\centering
	\includegraphics[width=\textwidth]{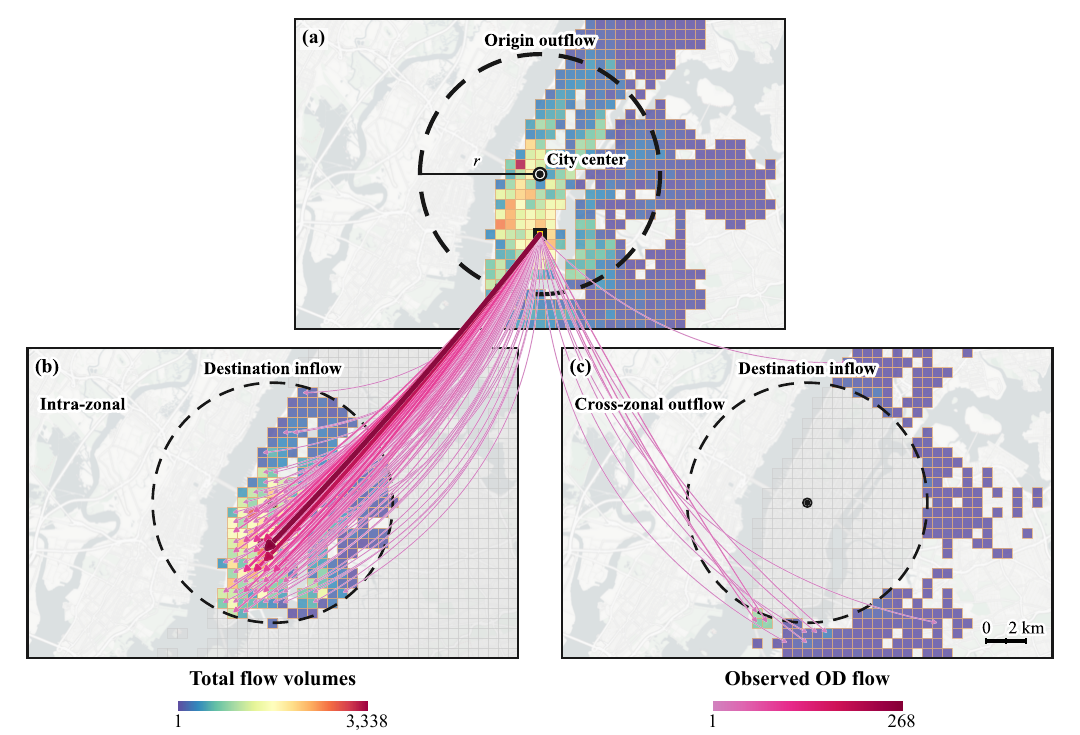}
	\caption{Spatial classification of cumulative weekday bike-sharing flows in New York City at \(L=500\)~m.
		(a) Trips originating inside a circular observation area of radius
		\(r\) centered on the city center.
		(b) Intra-zonal trips (II), with both endpoints inside the
		observation area.
		(c) Cross-zonal outflow (IO), with the origin inside and the
		destination outside.
		Grid colors indicate cumulative origin outflow in (a) and
		cumulative destination inflow in (b) and (c).
		Arrows show observed OD flows from the highlighted origin grid cell;
		arrow color and width indicate OD flow magnitude.}
	\label{fig:methods_spatial_modeling}
\end{figure}

\begin{figure}[h]
	\centering
	\includegraphics[width=\textwidth]{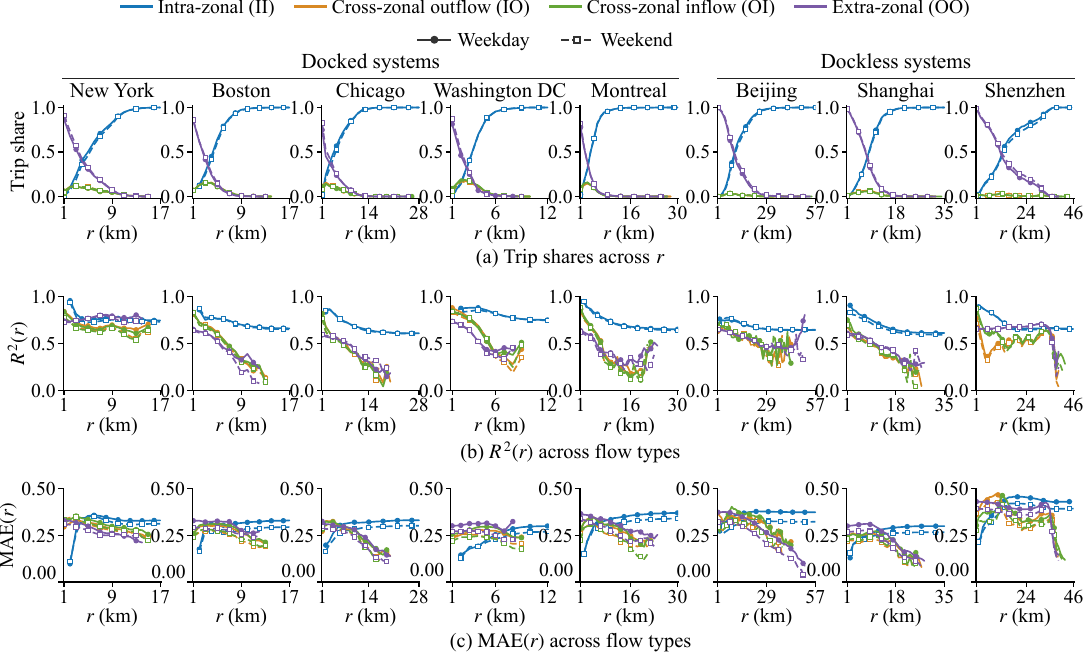}
	\caption{Spatial trip composition and gravity model performance across observation radius \(r\). (a) Shares of intra-zonal trips (II), cross-zonal outflow (IO), cross-zonal inflow (OI), and extra-zonal trips (OO); (b) \(R^2(r)\) for II, IO, OI, and OO; (c) \(\mathrm{MAE}(r)\) for II, IO, OI, and OO. }
	\label{fig:spatial_fit}
\end{figure}

The following spatial analysis uses a grid-cell size of \(L=1000\)~m
and expands the circular observation area from the city center by
increasing its radius \(r\). Figure~\ref{fig:methods_spatial_modeling} illustrates the expanding
observation area and the II and IO flow categories.
 Figure~\ref{fig:spatial_fit} shows that trip composition and gravity model performance vary with the observation radius \(r\), and that these changes differ among the four flow categories. At small radii, most trips are OO because both endpoints are outside the circular observation area. As \(r\) increases, the OO share decreases and the II share increases toward one (Figure~\ref{fig:spatial_fit}(a)). The IO and OI shares first increase, reach their highest values at intermediate radii, and then decrease as the observation area continues to expand. Their shares remain close across most radii, showing similar volumes of cross-zonal outflow and inflow.
Gravity model performance also changes with \(r\). For II, \(R^2(r)\) generally decreases and then remains relatively stable at larger radii, whereas \(\mathrm{MAE}(r)\) increases rapidly at small radii and subsequently changes little (Figures~\ref{fig:spatial_fit}(b) and \ref{fig:spatial_fit}(c)). The decrease in \(R^2(r)\) together with the increase in \(\mathrm{MAE}(r)\) shows weaker fit within the II flows observed at larger radii. For IO, OI, and OO, \(R^2(r)\) generally decreases but fluctuates more across radii and cities (Figure~\ref{fig:spatial_fit}(b)). Their \(\mathrm{MAE}(r)\) often reaches high values at small or intermediate radii and then decreases (Figure~\ref{fig:spatial_fit}(c)). Gravity model performance for these flow categories therefore changes less uniformly with \(r\). Weekday and weekend curves generally show the same radial patterns.

\begin{figure}[h]
	\centering
	\includegraphics[width=0.98\textwidth]{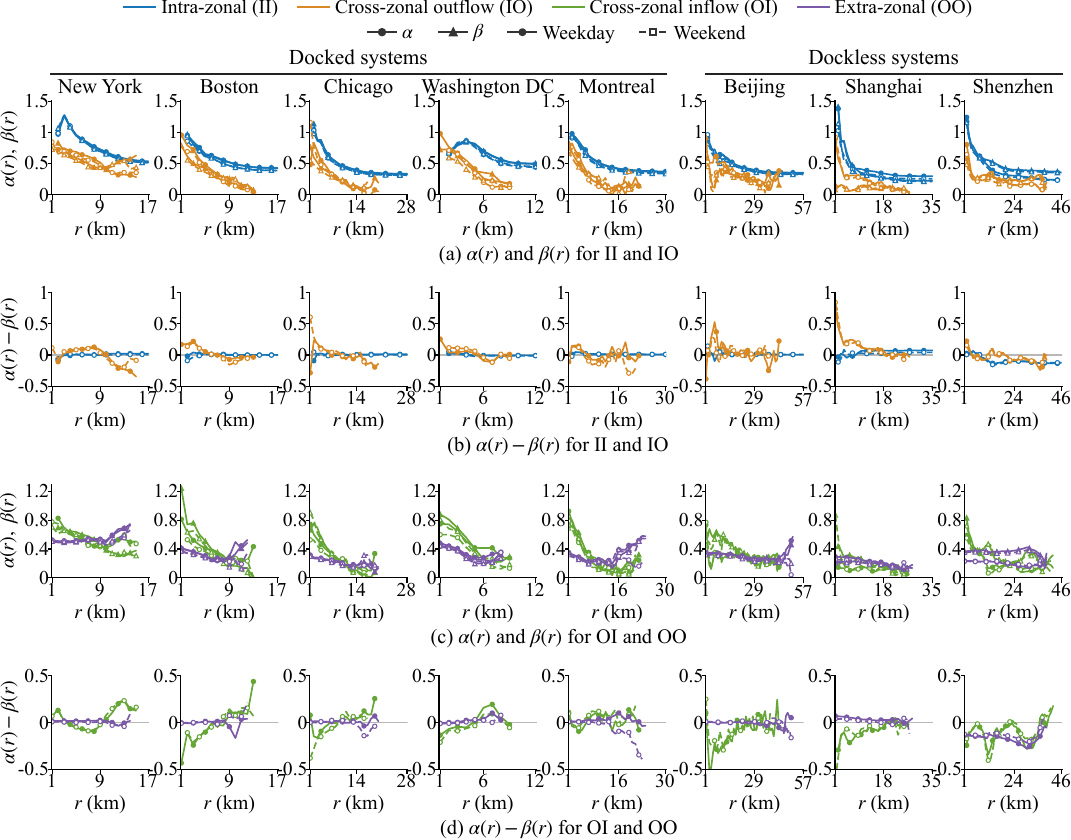}
	\caption{Origin and destination dependence across observation radius \(r\). (a) \(\alpha(r)\) and \(\beta(r)\) for II and IO; (b) \(\alpha(r)-\beta(r)\) for II and IO; (c) \(\alpha(r)\) and \(\beta(r)\) for OI and OO; (d) \(\alpha(r)-\beta(r)\) for OI and OO.}
	\label{fig:spatial_parameters}
\end{figure}

Figure~\ref{fig:spatial_parameters} shows that origin dependence and destination dependence change differently among the four flow categories. For II, \(\alpha(r)\) and \(\beta(r)\) generally decrease from approximately \(1\) at small radii to \(0.3\)--\(0.4\) at the outer radii and remain close across the observation range (Figure~\ref{fig:spatial_parameters}(a)). Origin and destination dependence both decrease as the circular observation area expands. Accordingly, \(\alpha(r)-\beta(r)\) remains near zero in most cities (Figure~\ref{fig:spatial_parameters}(b)). Origin out-strength and destination in-strength therefore have similar effects on intra-zonal flows in most cities. Shenzhen is the main exception, with \(\alpha(r)-\beta(r)\) remaining negative across most radii, indicating stronger destination dependence.

For IO, both \(\alpha(r)\) and \(\beta(r)\) generally decrease with \(r\), while \(\alpha(r)-\beta(r)\) decreases from positive or near-zero values toward negative values in most cities (Figures~\ref{fig:spatial_parameters}(a) and \ref{fig:spatial_parameters}(b)). Destination dependence therefore becomes stronger relative to origin dependence as the observation area expands. OI shows the opposite change: \(\alpha(r)-\beta(r)\) generally increases from negative values toward zero or positive values, indicating increasing origin dependence relative to destination dependence (Figures~\ref{fig:spatial_parameters}(c) and \ref{fig:spatial_parameters}(d)). For OO, \(\alpha(r)\), \(\beta(r)\), and their difference vary more among cities across radii. These broad patterns appear on both weekdays and weekends.

\begin{figure}[H]
	\centering
	\includegraphics[width=0.98\textwidth]{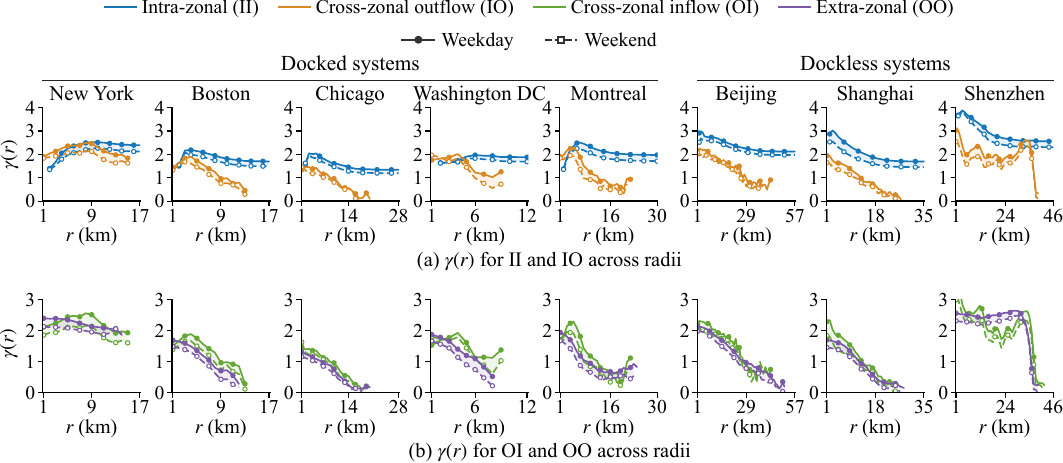}
	\caption{Distance dependence across observation radius \(r\). (a) \(\gamma(r)\) for II and IO; (b) \(\gamma(r)\) for OI and OO.}
	\label{fig:spatial_distance_decay}
\end{figure}

Figure~\ref{fig:spatial_distance_decay} shows clear differences in the distance dependence exponent \(\gamma(r)\) across observation radii. Weekday \(\gamma(r)\) generally exceeds weekend \(\gamma(r)\) across the four flow categories, reflecting stronger distance dependence on weekdays. Across observation radii, II \(\gamma(r)\) ranges from approximately \(1.2\) to \(2.5\) in the docked systems and from \(1.5\) to \(3.0\) in Beijing and Shanghai, whereas Shenzhen shows the strongest distance dependence for II, with \(\gamma(r)\) ranging from \(2.3\) to \(3.9\) (Figure~\ref{fig:spatial_distance_decay}(a)). Its change with \(r\) differs among cities: it increases at small radii and then changes little in New York, remains relatively stable in Washington DC, and generally decreases after the small-radius range in the other cities.

For IO, OI, and OO, \(\gamma(r)\) generally decreases as \(r\) increases, although Shenzhen shows less regular changes (Figures~\ref{fig:spatial_distance_decay}(a) and \ref{fig:spatial_distance_decay}(b)). This decrease reflects weaker distance dependence at larger observation radii.

The results obtained using the cycling activity centers are generally consistent with those obtained using the city centers, indicating that the choice of center has little effect on the overall radial patterns, although the exact values and local fluctuations differ across some cities and radii (see Supplementary Section~S6). Joint spatiotemporal results for weekdays and weekends are reported in Supplementary Sections~S7 and~S8, respectively.

\section{Discussion}
\label{sec:conclusion}

Gravity-model parameters are often treated as characteristics of a city or mobility system, although they are estimated from flows observed over a particular period and spatial extent. Across five docked and three dockless bike-sharing systems, we find that origin, destination and distance exponents, together with model performance, vary systematically within the day and with observation radius. These changes remain when grid resolution is fixed and the same full-day origin out-strengths and destination in-strengths are used across time windows and radii. Full-day and full-area estimates therefore summarize different temporal states and spatially defined flow subsets rather than a single stable relationship between mobility flows, activity and distance.

Previous studies showed that gravity-model parameters vary with spatial resolution and across weekdays, weekends, seasons and longer observation periods \cite{LiR2021,Jouve2025}, while aggregate mobility flows can often be represented by parsimonious gravity-like models \cite{CabanasTirapu2025GravityLike}. Our analysis extends these findings by separating intraday variation from longer temporal differences and observation range from grid resolution. Sliding windows reveal changes within a day, whereas expanding circular observation areas reveal changes associated with spatial extent and boundary-defined flow composition.

Because origin out-strength and destination in-strength remain fixed across windows, changes in \(\alpha(t)\) and \(\beta(t)\) cannot be attributed to redefining the activity variables at each time of day. Instead, OD flows active in different windows relate differently to the same full-day activity structure. Destination dependence is stronger during weekday mornings in all eight systems, whereas origin dependence becomes stronger later in most cities. This pattern indicates a shift in the relative association between flows and their two endpoints, but it does not identify trip purpose, land use or the behavioral processes responsible for the shift. The weaker weekend contrast and persistent destination dependence in Shenzhen also show that the timing and magnitude of this shift remain city-specific.

\(R^2(t)\) and \(\mathrm{MAE}(t)\) provide complementary information. \(R^2(t)\) is generally lower in the early morning and higher during the daytime, whereas weekday \(\mathrm{MAE}(t)\) rises during the morning peak and then declines. Model performance therefore does not simply follow trip volume, and the two metrics should not be interpreted interchangeably.

Increasing the observation radius does more than enlarge the mapped area. It changes the OD pairs assigned to II, IO, OI and OO, together with their distance and activity distributions. For II flows, \(\alpha(r)\) and \(\beta(r)\) generally remain close, indicating similar origin and destination dependence among trips whose endpoints both lie inside the boundary. By contrast, \(\alpha(r)-\beta(r)\) changes in opposite directions for IO and OI in most cities. Because the endpoint inside the observation area is the origin for IO but the destination for OI, this contrast shows that the inferred origin--destination asymmetry depends on how the boundary partitions the flow system. The observation boundary is therefore part of the model specification rather than a neutral device for selecting trips.

Distance dependence is affected in a similar way. The fitted distance exponent is generally larger during active weekday periods than at night or on weekends, while \(\gamma(r)\) commonly decreases with radius for IO, OI and OO. These estimates are obtained from different temporally and spatially conditioned sets of positive OD flows. A full-day, full-area estimate may therefore conceal differences in how strongly distinct flow subsets are associated with distance.

The main temporal and radial patterns occur in both docked and dockless systems and remain broadly similar when the geographical city center is replaced by the cycling activity center. However, exact values and local fluctuations differ among cities. Differences in gravity-model parameters across cities or studies may therefore reflect both genuine mobility differences and differences in temporal aggregation, spatial resolution, observation range or boundary treatment. Comparative studies should harmonize these settings or explicitly evaluate sensitivity to them.

The datasets cover different years, observation periods and sample sizes, so they are better suited to identifying repeated qualitative patterns than to comparing absolute parameter values. Regular grids, Euclidean distances and circular observation areas also simplify actual transport networks and urban form. In addition, origin out-strength and destination in-strength are derived from the observed bike-sharing flows rather than independent external characteristics. The fitted exponents should therefore be interpreted as descriptive associations within this model, not as causal effects of urban activity or distance. Within these limits, the results show that the temporal and spatial support of observed OD flows is part of the meaning of a fitted gravity-model parameter.

\section{Methods}
\label{sec:methods}

\subsection{Data sources and study areas}
\label{sec:data_sources}
\label{sec:data}

This study analyzes bike-sharing trips from eight datasets in eight cities, covering two system types: docked systems in New York City (NYC), Boston, Chicago, Washington DC, and Montreal, and dockless systems in Beijing, Shanghai, and Shenzhen. Trip records for the five docked systems were obtained from Citi Bike, Bluebikes, Divvy, Capital Bikeshare, and BIXI, respectively. For the dockless systems, records for Beijing and Shanghai were obtained from Mobike, and those for Shenzhen from the Shenzhen Municipal Government Open Data Platform. Table~\ref{tab:data} reports the observation periods and descriptive statistics. Because the observation periods and sample sizes differ among systems, comparisons of descriptive trip patterns focus on their temporal and spatial distributions rather than absolute trip volumes. Weekdays are defined as Monday to Friday and weekends as Saturday and Sunday.

\begin{table}[H]
	\centering
	\caption{Data sources, observation periods, and descriptive statistics after filtering. Daily trips are averaged over the observed days. Distance and duration are trip-level means. Maximum distance denotes the distance from the city center to the farthest trip origin or destination, rounded up to the nearest kilometer. Beijing records do not contain trip end times, so trip duration cannot be calculated.}
	\label{tab:data}
	\footnotesize
	\setlength{\tabcolsep}{2.4pt}
	\renewcommand{\arraystretch}{1.12}
	\sisetup{
		detect-all,
		group-digits=all,
		group-separator={,},
		table-number-alignment=center
	}
	\begin{tabular*}{\textwidth}{@{\extracolsep{\fill}} l l l
			S[table-format=2.0]
			S[table-format=7.0]
			S[table-format=1.2]
			S[table-format=1.3]
			S[table-format=2.0]}
		\toprule
		City & Data source & Period
		& \multicolumn{1}{c}{Days}
		& \multicolumn{1}{c}{\shortstack{Mean daily\\trips}}
		& \multicolumn{1}{c}{\shortstack{Mean trip\\distance (km)}}
		& \multicolumn{1}{c}{\shortstack{Mean trip\\duration (h)}}
		& \multicolumn{1}{c}{\shortstack{Maximum distance\\from city center (km)}}
		\\
		\midrule
		\multicolumn{8}{l}{\textbf{Docked systems}}\\
		\midrule
		NYC           & Citi Bike         & Aug 2025 & 31 & 156037  & 2.19 & 0.227 & 17 \\
		Boston        & Bluebikes         & Aug 2025 & 31 & 16790   & 2.11 & 0.265 & 17 \\
		Chicago       & Divvy             & Aug 2025 & 31 & 16241   & 2.31 & 0.292 & 28 \\
		Washington DC & Capital Bikeshare & Aug 2025 & 31 & 13360   & 1.87 & 0.233 & 12 \\
		Montreal      & BIXI              & Aug 2025 & 31 & 72463   & 1.95 & 0.247 & 30 \\
		\midrule
		\multicolumn{8}{l}{\textbf{Dockless systems}}\\
		\midrule
		Beijing  & Mobike & May 2017 & 14 & 229447  & 0.81 & {--}  & 57 \\
		Shanghai & Mobike & Aug 2016 & 31 & 32786   & 1.56 & 0.280 & 35 \\
		Shenzhen & \shortstack[l]{Shenzhen Open\\Data Platform}
		& Aug 2021 & 31 & 1150016 & 1.08 & 0.214 & 46 \\
		\bottomrule
	\end{tabular*}
\end{table}

Administrative boundaries were used to define the study areas of NYC, Chicago, Washington DC, Beijing, Shanghai, and Shenzhen. The Boston study area was defined as the union of municipalities served by Bluebikes. The remote Salem service area was excluded, and the remaining boundary was extended outward by 1.5~km to retain nearby stations. The Montreal study area was defined as the union of municipalities served by BIXI. A trip was retained only if both its origin and destination fell within the corresponding study area.

The city center is used as the reference point for the spatial analysis of each city. The maximum distance in Table~\ref{tab:data} is the distance from the city center to the farthest trip origin or destination, rounded up to the nearest kilometer. The city centers used in the spatial analysis are listed in Supplementary Section~S5.
\subsection{Preprocessing and spatial aggregation}
\label{sec:preprocessing}

Trip record formats differ across systems. For the docked systems, records contain trip start and end times and origin and destination station identifiers. Station identifiers were mapped to the coordinates released with each dataset, and conventional bicycle and e-bike trips were combined. For the dockless systems, the Shanghai and Shenzhen records provide origin and destination coordinates together with trip start and end times. The Beijing records provide origin and destination geohashes and trip start times but no end times. Each geohash was decoded to the coordinates of the corresponding cell center. All origins and destinations were represented by geographic coordinates, and all times were expressed in local time.

Trips with a Euclidean distance between origin and destination longer than 20~km were discarded. For the seven datasets containing both start and end times, trips shorter than 1~min, longer than 12~h, or with an average speed faster than \(30~\mathrm{km}/\mathrm{h}\) were also discarded. Beijing records do not contain end times, so trip duration and speed could not be calculated.

Each study area was divided into regular square grid cells with side length \(L\). Trip origins and destinations were assigned to grid cells according to their coordinates and aggregated into OD flows between cells. Trips with both endpoints in the same cell were excluded from the intercell OD flows.

\subsection{Temporal modeling}
\label{sec:methods_temporal}

To describe changes in bike-sharing OD flows within a day, we define a sequence of observation start times \(t_0,t_1,t_2,\ldots,t_k,\ldots,t_n\). In this study, \(t_0=00{:}00\), and consecutive observation start times are separated by 1~h, so \(t_1=01{:}00\), \(t_2=02{:}00\), and so on until \(t_n=22{:}00\). The main observation window has length \(\tau=2\)~h, which retains intraday variation while providing sufficient positive-flow OD pairs for model estimation. We examine OD flows in the sliding windows \([t_0,t_0+\tau]\), \([t_1,t_1+\tau]\), \([t_2,t_2+\tau]\), \(\ldots\), \([t_k,t_k+\tau]\), \(\ldots\), and \([t_n,t_n+\tau]\) (Figure~\ref{fig:methods_windows}(a)). These windows correspond to 00:00--02:00, 01:00--03:00, and so on until 22:00--24:00, without crossing into the next day. Figure~\ref{fig:methods_windows}(b) shows the spatial distributions of origin trip volume on weekdays and weekends during \([t_8,t_8+\tau]\), \([t_{13},t_{13}+\tau]\), and \([t_{18},t_{18}+\tau]\), corresponding to 08:00--10:00, 13:00--15:00, and 18:00--20:00. Two alternative sliding-window settings are also considered:
(1) a 1~h observation window with a 0.5~h interval, corresponding
to 00:00--01:00, 00:30--01:30, and so on until 23:00--24:00; and
(2) a 3~h observation window with a 1~h interval, corresponding to
00:00--03:00, 01:00--04:00, and so on until 21:00--24:00
(see Supplementary Section~S4).

For any two grid cells \(i\) and \(j\), the flow-based gravity model
for the time window \([t_k,t_k+\tau]\) is
\begin{equation}
	T_{ij}^{[t_k,t_k+\tau]}
	=
	C_k
	\frac{O_i^{\alpha_k}D_j^{\beta_k}}
	{d_{ij}^{\gamma_k}},
	\qquad i\neq j.
	\label{eq:temporal_gravity}
\end{equation}

The model is estimated separately for each city, day type, and time
window after pooling trips from all observed days of the corresponding
day type. Here, \(T_{ij}^{[t_k,t_k+\tau]}\) is the cumulative flow
from grid cell \(i\) to grid cell \(j\) during the corresponding time
window across all observed days of that day type, \(C_k\) is a
proportionality constant, and \(d_{ij}\) is the Euclidean distance
between the grid centroids. For each city and day type,
\(O_i=\sum_{j\ne i}T_{ij}\) denotes the cumulative outflow from grid
cell \(i\) over all times and all observed days of that day type, and
\(D_j=\sum_{i\ne j}T_{ij}\) denotes the corresponding cumulative
inflow to grid cell \(j\), where \(T_{ij}\) is the cumulative flow
from grid cell \(i\) to grid cell \(j\) over the same pooled data. 

The same \(O_i\) and \(D_j\) are used across all time windows, providing a
fixed reference for comparing the fitted exponents over time. The
parameters \(\alpha_k\), \(\beta_k\), and \(\gamma_k\) measure origin,
destination, and distance dependence, respectively, in time window
\(k\). Larger values of \(\alpha_k\), \(\beta_k\), and \(\gamma_k\)
indicate stronger origin, destination, and distance dependence,
respectively. \(t\) denotes the start time \(t_k\) of the
corresponding observation window, so \(\alpha(t)\), \(\beta(t)\), and
\(\gamma(t)\) denote the plotted forms of \(\alpha_k\), \(\beta_k\),
and \(\gamma_k\), respectively.

\subsection{Spatial modeling}
\label{sec:methods_spatial}

For each city, a circular observation area with a specified radius \(r\) is constructed around the city center. For each center, \(r\) is increased from 1~km to the distance from that center to the farthest retained trip endpoint, rounded up to the
nearest kilometer, in 1~km increments. Trips originating inside the circular observation area are classified according to the location of their destination. Trips with both endpoints inside are classified as intra-zonal (II), whereas trips with the origin inside and the destination outside are classified as cross-zonal outflow (IO). Figure~\ref{fig:methods_spatial_modeling} illustrates these two categories. The spatial analysis also includes cross-zonal inflow (OI), with the origin outside and the destination inside the circular observation area, and extra-zonal trips (OO), with both endpoints outside. 

For any two grid cells \(i\) and \(j\), the flow-based gravity model
for a given radius \(r\) is
\begin{equation}
	T_{ij}(r)
	=
	C(r)
	\frac{O_i^{\alpha(r)}D_j^{\beta(r)}}
	{d_{ij}^{\gamma(r)}},
	\qquad i\neq j.
	\label{eq:spatial_gravity}
\end{equation}

The model is estimated separately for each city, day type, flow
category, and radius \(r\) after pooling trips from all observed days
of the corresponding day type. Here, \(T_{ij}(r)\) is the cumulative
flow from grid cell \(i\) to grid cell \(j\) in the corresponding flow
category across all observed days of that day type, \(C(r)\) is a
proportionality constant, and \(d_{ij}\) is the Euclidean distance
between the centroids of grid cells \(i\) and \(j\).

The same \(O_i\) and \(D_j\) are used across all
observation radii, providing a fixed reference for comparing the
fitted exponents across radii. The parameters \(\alpha(r)\),
\(\beta(r)\), and \(\gamma(r)\) measure origin, destination, and
distance dependence, respectively, at radius \(r\). The share of each
flow category is its total flow divided by the citywide total flow for
the corresponding day type. In addition to the city centers, we repeated the spatial analysis
using a cycling activity center, defined as the 1000~m grid cell with
the largest combined full-day inflow and outflow after pooling weekday
and weekend trips. The same cycling activity center is used for both
day types. The center definitions and results are reported in
Supplementary Sections~S5 and~S6.

\subsection{Reference models}
\label{sec:methods_reference}

Two full-day model specifications are considered for comparison.
When the full-day window \([t_0,t_n+\tau]\) and the complete study
area are used, the flow-based gravity model reduces to
\begin{equation}
	T_{ij}
	=
	C
	\frac{O_i^{\alpha}D_j^{\beta}}
	{d_{ij}^{\gamma}},
	\qquad i\neq j.
	\label{eq:full_domain_gravity}
\end{equation}

Here, \(T_{ij}\) is the cumulative full-day OD flow from grid cell
\(i\) to grid cell \(j\) across all observed days of the corresponding
day type, \(C\) is a proportionality constant, and \(d_{ij}\) is the
Euclidean distance between the grid centroids. \(O_i\) and \(D_j\)
are the corresponding cumulative full-day total outflow from grid
cell \(i\) and total inflow to grid cell \(j\), respectively. The
parameters \(\alpha\), \(\beta\), and \(\gamma\) measure origin,
destination, and distance dependence, respectively.

Replacing \(O_i\) and \(D_j\) with the population sizes \(m_i\) and
\(m_j\) gives the population-based gravity model used in previous
bike-sharing research \cite{LiR2021}:
\begin{equation}
	T_{ij}
	=
	C
	\frac{m_i^{\alpha}m_j^{\beta}}
	{d_{ij}^{\gamma}},
	\qquad i\neq j.
	\label{eq:population_gravity}
\end{equation}

For each city, \(m_i\) and \(m_j\) are obtained by aggregating the
WorldPop raster for the year of the corresponding trip data from its
native resolution of approximately 100~m to the analysis grid
\cite{Tatem2017}.

\subsection{Estimation and evaluation}
\label{sec:methods_estimation}
The gravity model parameters are estimated by ordinary least squares
in \(\log_{10}\) space using OD pairs with positive observed flows and
\(i \ne j\). Model performance is evaluated using the mean absolute error (MAE) and the coefficient of determination (\(R^2\)) in \(\log_{10}\) space:
\begin{equation}
	\mathrm{MAE}
	=
	\frac{1}{N}
	\sum_{i,j,\;i\neq j,\;T_{ij}>0}
	\left|
	\log_{10}\!\left(\hat{T}_{ij}\right)
	-
	\log_{10}\!\left(T_{ij}\right)
	\right|,
	\label{eq:mae}
\end{equation}
\begin{equation}
	R^2
	=
	1-
	\frac{
		\sum_{i,j,\;i\neq j,\;T_{ij}>0}
		\left[
		\log_{10}\!\left(\hat{T}_{ij}\right)
		-
		\log_{10}\!\left(T_{ij}\right)
		\right]^2
	}{
		\sum_{i,j,\;i\neq j,\;T_{ij}>0}
		\left[
		\log_{10}\!\left(T_{ij}\right)
		-
		\frac{1}{N}
		\sum_{i,j,\;i\neq j,\;T_{ij}>0}
		\log_{10}\!\left(T_{ij}\right)
		\right]^2
	}.
	\label{eq:r2}
\end{equation}

Here, \(N\) is the number of fitted OD pairs with \(i\neq j\) and \(T_{ij}>0\), \(\hat{T}_{ij}\) denotes the fitted flow, and \(\frac{1}{N}\sum_{i,j,\;i\neq j,\;T_{ij}>0}\allowbreak\log_{10}\!\left(T_{ij}\right)\) is the mean of the observed flows on the log scale. To avoid unstable estimates in sparse temporal, spatial, and spatiotemporal subsets, results are reported only when at least 40 positive-flow OD pairs are available. Results involving \(\alpha\) or \(\beta\) are additionally excluded when the standard error of either exponent is greater than 0.30. 

The two metrics describe complementary aspects of model performance. \(R^2\) measures the proportion of variation in observed log flows explained by the model relative to the mean observed log flow. \(\mathrm{MAE}\) measures the mean absolute difference between fitted and observed log flows across individual OD pairs. Because both the positive-flow observations and the variation in observed log flows differ among time windows, observation radii, and flow categories, \(R^2\) and \(\mathrm{MAE}\) describe model fit within each corresponding subset. We also examined how the performance of the flow-based and
population-based gravity models varies with trip distance on weekdays;
the analysis is described in Supplementary Section~S2.
	
\section*{Data availability}
	Trip records for New York City, Boston, Chicago, Washington DC,
		and Montreal are publicly available from the Citi Bike System Data
		page (\url{https://citibikenyc.com/system-data}), the Bluebikes System
		Data page (\url{https://bluebikes.com/system-data}), the Divvy Data
		page (\url{https://divvybikes.com/system-data}), the Capital Bikeshare
		System Data page
		(\url{https://capitalbikeshare.com/system-data}), and the BIXI
		Montr\'eal Open Data page (\url{https://bixi.com/en/open-data}),
		respectively. The Shenzhen trip records can be accessed through the bike-sharing daily-order data API provided by the Shenzhen Municipal Government Open Data Platform
		(\url{https://opendata.sz.gov.cn/data/api/toApiDetails/29200_00403627}).
		The Mobike trip records for Beijing and Shanghai were obtained from a
		third party and are not publicly available.

\section*{Code availability}
The code supporting the findings of this study is available from the corresponding author upon reasonable request.

\section*{Acknowledgements}
	This work was supported by the Yunnan Fundamental Research Projects (Grant No. 202401AT070359).

\section*{Author contributions}
\textbf{Hanbo Zhang}: Conceptualization, Methodology, Formal analysis, Data curation, Software, Writing -- original draft.
\textbf{Qi Rao}: Data curation, Visualization, Writing -- review \& editing.
\textbf{Bo Yang}: Conceptualization, Methodology, Formal analysis, Funding acquisition, Writing -- review \& editing.

\section*{Competing interests}
	The authors declare no competing interests.

	\section*{Supplementary Information}
		Supplementary Information for this article contains descriptive statistics,
		additional model comparisons, robustness tests across spatial resolutions and temporal windows,
		city-center definitions, spatial analyses based on cycling activity centers,
		and weekday and weekend spatiotemporal results.

	\bibliographystyle{elsarticle-num}
	\bibliography{refs}

@article{Zhao2024,
	author  = {Zhao, Pengjun and Wang, Hao and Liu, Qiyang and Yan, Xiao-Yong and Li, Jingzhong},
	title   = {Unravelling the spatial directionality of urban mobility},
	journal = {Nat. Commun.},
	year    = {2024},
	volume  = {15},
	pages   = {4507},
	doi     = {10.1038/s41467-024-48909-7}
}

@article{Zhong2025Universal,
	author  = {Zhong, Lu and Dong, Lei and Wang, Qi R. and Song, Chaoming and Gao, Jianxi},
	title   = {Universal Expansion of Human Mobility across Urban Scales},
	journal = {Nat. Cities},
	volume  = {2},
	pages   = {603--607},
	year    = {2025},
	doi     = {10.1038/s44284-025-00268-0}
}

@article{Xu2021UrbanGrowth,
	author  = {Xu, Fengli and Li, Yong and Jin, Depeng and Lu, Jianhua and Song, Chaoming},
	title   = {Emergence of urban growth patterns from human mobility behavior},
	journal = {Nat. Comput. Sci.},
	volume  = {1},
	pages   = {791--800},
	year    = {2021},
	doi     = {10.1038/s43588-021-00160-6}
}

@article{Ramani2024WFH,
	author  = {Ramani, Arjun and Alcedo, Joel and Bloom, Nicholas},
	title   = {How working from home reshapes cities},
	journal = {Proc. Natl Acad. Sci. USA},
	volume  = {121},
	number  = {45},
	pages   = {e2408930121},
	year    = {2024},
	doi     = {10.1073/pnas.2408930121}
}

@article{Winkler2023SustainableMobility,
	author  = {Winkler, Lisa and Pearce, Drew and Nelson, Jenny and Babacan, Oytun},
	title   = {The effect of sustainable mobility transition policies on cumulative urban transport emissions and energy demand},
	journal = {Nat. Commun.},
	volume  = {14},
	pages   = {2357},
	year    = {2023},
	doi     = {10.1038/s41467-023-37728-x}
}

@article{Asensio2022Micromobility,
	author  = {Asensio, Omar Isaac and Apablaza, Camila Z. and Lawson, M. Cade and Chen, Edward W. and Horner, Savannah J.},
	title   = {Impacts of micromobility on car displacement with evidence from a natural experiment and geofencing policy},
	journal = {Nat. Energy},
	volume  = {7},
	pages   = {1100--1108},
	year    = {2022},
	doi     = {10.1038/s41560-022-01135-1}
}

@article{Liang2024SharedMobilityOD,
	author  = {Liang, Yuebing and Zhao, Zhan and Webster, Chris},
	title   = {Generating sparse origin--destination flows on shared mobility networks using probabilistic graph neural networks},
	journal = {Sustain. Cities Soc.},
	volume  = {114},
	pages   = {105777},
	year    = {2024},
	doi     = {10.1016/j.scs.2024.105777}
}

@article{Xu2023JAG,
	author  = {Xu, Xijie and Wang, Jie and Poslad, Stefan and Rui, Xiaoping and Zhang, Guangyuan and Fan, Yonglei},
	title   = {Exploring intra-urban human mobility and daily activity patterns from the lens of dockless bike-sharing: A case study of {Beijing}, {China}},
	journal = {Int. J. Appl. Earth Obs. Geoinf.},
	year    = {2023},
	volume  = {122},
	pages   = {103442},
	doi     = {10.1016/j.jag.2023.103442}
}

@article{Meng2023,
	author  = {Meng, Fanyun and Zheng, Lili and Ding, Tongqiang and Wang, Zhuorui and Zhang, Yanlin and Li, Wenqing},
	title   = {Understanding dockless bike-sharing spatiotemporal travel patterns: Evidence from ten cities in {China}},
	journal = {Comput. Environ. Urban Syst.},
	volume  = {104},
	pages   = {102006},
	year    = {2023},
	doi     = {10.1016/j.compenvurbsys.2023.102006}
}

@article{Tan2025Spatiotemporal,
	author  = {Tan, Xingye and Huang, Bo and Batty, Michael and Li, Weiyu and Wang, Qi Ryan and Zhou, Yulun and Gong, Peng},
	title   = {The Spatiotemporal Scaling Laws of Urban Population Dynamics},
	journal = {Nat. Commun.},
	volume  = {16},
	pages   = {2881},
	year    = {2025},
	doi     = {10.1038/s41467-025-58286-4}
}

@article{Abbiasov2024FifteenMinute,
	author  = {Abbiasov, Timur and Heine, Cate and Sabouri, Sadegh and Salazar-Miranda, Arianna and Santi, Paolo and Glaeser, Edward and Ratti, Carlo},
	title   = {The 15-Minute City Quantified Using Human Mobility Data},
	journal = {Nat. Hum. Behav.},
	volume  = {8},
	pages   = {445--455},
	year    = {2024},
	doi     = {10.1038/s41562-023-01770-y}
}

@article{Barbosa2018,
	author  = {Barbosa, Hugo and Barthelemy, Marc and Ghoshal, Gourab and James, Charlotte R. and Lenormand, Maxime and Louail, Thomas and Menezes, Ronaldo and Ramasco, Jos{\'e} J. and Simini, Filippo and Tomasini, Marcello},
	title   = {Human mobility: Models and applications},
	journal = {Phys. Rep.},
	year    = {2018},
	volume  = {734},
	pages   = {1--74},
	doi     = {10.1016/j.physrep.2018.01.001}
}

@article{Zipf1946,
	author  = {Zipf, George K.},
	title   = {The {P1 P2/D} hypothesis: On the intercity movement of persons},
	journal = {Am. Sociol. Rev.},
	year    = {1946},
	volume  = {11},
	number  = {6},
	pages   = {677--686},
	doi     = {10.2307/2087063}
}

@article{Wilson1967TR,
	author  = {Wilson, Alan G.},
	title   = {A statistical theory of spatial distribution models},
	journal = {Transp. Res.},
	year    = {1967},
	volume  = {1},
	number  = {3},
	pages   = {253--269},
	doi     = {10.1016/0041-1647(67)90035-4}
}

@article{Evans1973TR,
	author  = {Evans, Suzanne P.},
	title   = {A relationship between the gravity model for trip distribution and the transportation problem in linear programming},
	journal = {Transp. Res.},
	year    = {1973},
	volume  = {7},
	number  = {1},
	pages   = {39--61},
	doi     = {10.1016/0041-1647(73)90005-1}
}

@article{Jung2008,
	author  = {Jung, Woo-Sung and Wang, Fengzhong and Stanley, H. Eugene},
	title   = {Gravity model in the {Korean} highway},
	journal = {EPL},
	year    = {2008},
	volume  = {81},
	pages   = {48005},
	doi     = {10.1209/0295-5075/81/48005}
}

@article{Goh2012,
	author  = {Goh, Segun and Lee, Keumsook and Park, Jong Soo and Choi, M. Y.},
	title   = {Modification of the gravity model and application to the metropolitan {Seoul} subway system},
	journal = {Phys. Rev. E},
	year    = {2012},
	volume  = {86},
	number  = {2},
	pages   = {026102},
	doi     = {10.1103/PhysRevE.86.026102}
}

@article{HongJung2016,
	author  = {Hong, Inho and Jung, Woo-Sung},
	title   = {Application of gravity model on the {Korean} urban bus network},
	journal = {Physica A},
	year    = {2016},
	volume  = {462},
	pages   = {48--55},
	doi     = {10.1016/j.physa.2016.06.055}
}

@article{Simini2012,
	author  = {Simini, Filippo and Gonz{\'a}lez, Marta C. and Maritan, Amos and Barab{\'a}si, Albert-L{\'a}szl{\'o}},
	title   = {A universal model for mobility and migration patterns},
	journal = {Nature},
	year    = {2012},
	volume  = {484},
	number  = {7392},
	pages   = {96--100},
	doi     = {10.1038/nature10856}
}

@article{LiuYan2020,
	author  = {Liu, Er-Jian and Yan, Xiao-Yong},
	title   = {A universal opportunity model for human mobility},
	journal = {Sci. Rep.},
	year    = {2020},
	volume  = {10},
	pages   = {4657},
	doi     = {10.1038/s41598-020-61613-y}
}

@article{Simini2021,
	author  = {Simini, Filippo and Barlacchi, Gianni and Luca, Massimiliano and Pappalardo, Luca},
	title   = {A Deep Gravity model for mobility flows generation},
	journal = {Nat. Commun.},
	year    = {2021},
	volume  = {12},
	number  = {1},
	pages   = {6576},
	doi     = {10.1038/s41467-021-26752-4}
}

@article{Yang2026NeuroGravity,
	author  = {Yang, Jinming and Huang, Shaoyu and Huang, Zongyuan and Jin, Yaohui and Yang, Xiaokang and Gonz{\'a}lez, Marta C. and Xu, Yanyan},
	title   = {Transferable human mobility network reconstruction with {neuroGravity}},
	journal = {Nat. Comput. Sci.},
	volume  = {6},
	pages   = {630--641},
	year    = {2026},
	doi     = {10.1038/s43588-026-01003-y}
}

@article{CabanasTirapu2025GravityLike,
	author  = {Cabanas-Tirapu, Oriol and Dan{\'u}s, Llu{\'i}s and Moro, Esteban and Sales-Pardo, Marta and Guimer{\`a}, Roger},
	title   = {Human mobility is well described by closed-form gravity-like models learned automatically from data},
	journal = {Nat. Commun.},
	year    = {2025},
	volume  = {16},
	pages   = {1336},
	doi     = {10.1038/s41467-025-56495-5}
}

@article{LiR2021,
	author  = {Li, Ruiqi and Gao, Shuai and Luo, Ankang and Yao, Qing and Chen, Bingsheng and Shang, Fan and Jiang, Rui and Stanley, H. Eugene},
	title   = {Gravity model in dockless bike-sharing systems within cities},
	journal = {Phys. Rev. E},
	year    = {2021},
	volume  = {103},
	number  = {1},
	pages   = {012312},
	doi     = {10.1103/PhysRevE.103.012312}
}

@article{Jouve2025,
	author  = {Jouve, Bertrand and Rochet, Paul and Salifou, Mohamadou},
	title   = {A comparative study of gravity models to assess the evolution of urban bicycle mobility},
	journal = {Trans. Transp. Sci.},
	year    = {2025},
	volume  = {16},
	number  = {1},
	pages   = {4--13},
	doi     = {10.5507/tots.2024.019}
}

@article{Tatem2017,
	author  = {Tatem, Andrew J.},
	title   = {{WorldPop}, open data for spatial demography},
	journal = {Sci. Data},
	year    = {2017},
	volume  = {4},
	pages   = {170004},
	doi     = {10.1038/sdata.2017.4}
}

@article{Waldner2025BikeSharing,
	author  = {Waldner, Felix and Balke, Georg and Rech, Felix and Lellep, Martin},
	title   = {Data-driven insights into ({E}-)bike-sharing: mining a large-scale dataset on usage and urban characteristics: descriptive analysis and performance modeling},
	journal = {Transportation},
	year    = {2025},
	volume  = {52},
	pages   = {2433--2473},
	doi     = {10.1007/s11116-025-10661-2}
}

@article{LiWang2025SharedBicycle,
	author  = {Li, Na and Wang, Tianqun},
	title   = {Analysis of the spatio-temporal impact of the built environment on shared bicycle ridership density},
	journal = {Comput. Urban Sci.},
	year    = {2025},
	volume  = {5},
	pages   = {1},
	doi     = {10.1007/s43762-024-00153-x}
}

@article{Lu2025BikeSharing,
	author  = {Lu, Kai-Fa and Liu, Yanghe and Peng, Zhong-Ren},
	title   = {Assessing the impacts of transit systems and urban street features on bike-sharing ridership: A graph-based spatiotemporal analysis and prediction model},
	journal = {J. Transp. Geogr.},
	year    = {2025},
	volume  = {128},
	pages   = {104356},
	doi     = {10.1016/j.jtrangeo.2025.104356}
}

@article{Pappalardo2023FutureDirections,
	author  = {Pappalardo, Luca and Manley, Ed and Sekara, Vedran and Alessandretti, Laura},
	title   = {Future directions in human mobility science},
	journal = {Nat. Comput. Sci.},
	year    = {2023},
	volume  = {3},
	number  = {7},
	pages   = {588--600},
	doi     = {10.1038/s43588-023-00469-4}
}

@article{Santana2023TimeSpace,
	author  = {Santana, Clodomir and Botta, Federico and Barbosa, Hugo and Privitera, Filippo and Menezes, Ronaldo and Di Clemente, Riccardo},
	title   = {{COVID-19} is linked to changes in the time--space dimension of human mobility},
	journal = {Nat. Hum. Behav.},
	year    = {2023},
	volume  = {7},
	number  = {10},
	pages   = {1729--1739},
	doi     = {10.1038/s41562-023-01660-3}
}

@article{Boucherie2025Geography,
	author  = {Boucherie, Louis and Maier, Benjamin F. and Lehmann, Sune},
	title   = {Decoupling geographical constraints from human mobility},
	journal = {Nat. Hum. Behav.},
	year    = {2025},
	volume  = {9},
	number  = {12},
	pages   = {2564--2575},
	doi     = {10.1038/s41562-025-02282-7}
}

@article{Grosche2007,
	author  = {Grosche, Tobias and Rothlauf, Franz and Heinzl, Armin},
	title   = {Gravity models for airline passenger volume estimation},
	journal = {J. Air Transp. Manag.},
	year    = {2007},
	volume  = {13},
	number  = {4},
	pages   = {175--183},
	doi     = {10.1016/j.jairtraman.2007.02.001}
}

@article{Masucci2013,
	author  = {Masucci, A. Paolo and Serras, Joan and Johansson, Anders and Batty, Michael},
	title   = {Gravity versus radiation models: On the importance of scale and heterogeneity in commuting flows},
	journal = {Phys. Rev. E},
	year    = {2013},
	volume  = {88},
	number  = {2},
	pages   = {022812},
	doi     = {10.1103/PhysRevE.88.022812}
}

@article{Kwon2023MultipleGravity,
	author  = {Kwon, Oh-Hyun and Hong, Inho and Jung, Woo-Sung and Jo, Hang-Hyun},
	title   = {Multiple gravity laws for human mobility within cities},
	journal = {EPJ Data Sci.},
	year    = {2023},
	volume  = {12},
	pages   = {57},
	doi     = {10.1140/epjds/s13688-023-00438-x}
}

@article{Thompson2019,
	author  = {Thompson, C. A. and Saxberg, K. and Lega, J. and Tong, D. and Brown, H. E.},
	title   = {A cumulative gravity model for inter-urban spatial interaction at different scales},
	journal = {J. Transp. Geogr.},
	year    = {2019},
	volume  = {79},
	pages   = {102461},
	doi     = {10.1016/j.jtrangeo.2019.102461}
}

@article{ZhangX2024,
	author  = {Zhang, Xinyuan and Li, Nan},
	title   = {An activity space-based gravity model for intracity human mobility flows},
	journal = {Sustain. Cities Soc.},
	year    = {2024},
	volume  = {101},
	pages   = {105073},
	doi     = {10.1016/j.scs.2023.105073}
}
	
	\end{document}